\documentclass[aps,pra, noeprint, longbibliography]{revtex4-2}

\usepackage{amsmath}
\usepackage{comment}
\usepackage{caption}
\usepackage{fixltx2e}
\usepackage{graphicx}
\usepackage{setspace}
\usepackage[margin=2.5cm]{geometry} 

\usepackage{textgreek}
\usepackage{gensymb}
\usepackage{enumitem}
\usepackage[colorlinks=true,
    linkcolor=blue,
    citecolor=blue,
    urlcolor=blue]{hyperref}

\begin{document}
\title[High-dimensional reinforcement learning for ultracold quantum gases]{High-dimensional reinforcement learning for optimization and control of ultracold quantum gases}
\author{Nicholas Milson$^1$}
\email[Corresponding author: ]{nmilson@ualberta.ca}
\author{Arina Tashchilina$^1$}
\author{Tian Ooi$^1$}
\author{Anna Czarnecka$^1$}
\author{Zaheen F.\ Ahmad$^2$}
\author{Lindsay J.\ LeBlanc$^1$}
\email{lindsay.leblanc@ualberta.ca}
\affiliation{$^1$Department of Physics, University of Alberta, Edmonton, Alberta, Canada}
\affiliation{$^2$Department of Computing Sciences, University of Alberta, Edmonton, Alberta, Canada}

\begin{abstract}
Machine-learning techniques are emerging as a valuable tool in experimental physics, and among them, reinforcement learning offers the  potential to control high-dimensional, multistage  processes in the presence of fluctuating environments. In this experimental work, we apply reinforcement learning to the preparation of an ultracold quantum gas to realize a consistent and large number of atoms at microkelvin temperatures. This reinforcement learning agent determines an optimal set of thirty control parameters in a dynamically changing environment that is characterized by thirty sensed parameters. By comparing this method to that of training supervised-learning regression models, as well as to human-driven control schemes, we find that both  machine learning approaches accurately predict the number of cooled atoms and both  result in occasional superhuman control schemes. However,  only the reinforcement learning method achieves consistent outcomes, even in the presence of a dynamic environment. 
\end{abstract}

\vspace{2pc}
\maketitle

\section{Introduction}

As a general tool, machine learning (ML) offers remarkable advantages in far-reaching domains, ranging from large-language models to control-systems instrumentation. In the realm of scientific research, ML promises to improve the design, control, and optimization of experimental processes, particularly when these procedures are high-dimensional, when uncontrolled environmental factors affect outcomes, and when systematic optimization is implausible, leaving only intuition or trial and error~\cite{krenn2020computer}. The production of cold, dense ensembles of neutral atoms for Bose-Einstein condensates (BECs), for example, involves many detailed steps, each with several free parameters~\cite{heck2018remote}.  While experimentalists have risen to the challenge to create systems that consistently result in BECs, applications of  ML to optimize magneto-optical traps \cite{tranter2018multiparameter,xu2021maximizing}, evaporation curves \cite{wigley2016fast,nakamura2019non,wu2020active,davletov2020machine,ma2023bayesian}, and simultaneous laser and evaporative cooling \cite{barker2020applying,vendeiro2022machine} show the potential for creating greater reliability for these systems.

These impressive results in atom-cooling applications are examples of supervised ML, in which a model learns from labeled examples, aiming to predict or classify new instances of the same problem based on prior training. So far, these approaches used Gaussian processes~\cite{seeger2004gaussian} or  deep neural networks~\cite{mehta2019high} to apply a supervised ML regression model (sometimes known as a surrogate function), then used the trained model to find experimental control inputs that maximize the predicted output.  
These methods to find efficient cooling schemes demonstrated superior performance compared to human optimization and direct numerical optimization techniques, such as differential evolution~\cite{wigley2016fast}. However, in scenarios where optimal control parameters depend on changing environmental conditions, these approaches are inadequate. A direct maximization  via Bayesian optimization~\cite{snoek2012practical} is unsuited, as the objective function varies with the environment. Furthermore, modeling the output metric as a function of controllable and environmental factors and optimizing the model alone is likely insufficient, as previous investigations that only considered the controllable factors failed to yield consistent superhuman control parameter choices~\cite{tranter2018multiparameter,barker2020applying}. Robustness of the found experimental control inputs is now also a consideration, as the usefulness of finding a single high-performing set is now highly dependent on how strongly the performance of this set changes with the dynamic environment.

Beyond the scope of supervised ML, machine learning also includes two more general approaches: unsupervised ML, which deals with finding patterns and structures in unlabeled data, through techniques such as clustering and dimensionality reduction; and reinforcement learning (RL), which represents a class of problems where an agent interacts with an environment, learning to make sequential decisions to maximize cumulative rewards. While still a young subfield in physics, the power of reinforcement learning has been demonstrated in the control of noisy and many-body quantum systems \cite{ding2023closed, bukov2018reinforcement, haug2021machine, cardenas2018multiqubit, lamata2017basic,seif2018machine}, as well as experimental control and navigation in stochastic turbulent environments such as  micro-sphere carrying optical tweezers \cite{praeger2021playing} and microorganism simulating artificial nano-swimmers \cite{colabrese2017flow}. Despite this recent wave of RL successes, live, autonomous optimization and control of atom-cooling apparatuses with RL remains largely unexplored, particularly in the context of high-dimensional control-parameter and environmental-parameter spaces.

In this paper, we optimize a rubidium-87 atom cooling experiment using  ML approaches that consider the environment the apparatus is in. We focus our work on the crucial initial production stages of a BEC, including laser cooling and trapping, up to a high-field-gradient magnetic trap (MT)~\cite{ketterle1999making,Lin2009-qz}. Our high-dimensional ML schemes include thirty measured environmental parameters, sensed at specific times throughout the experimental sequence, in addition to thirty control parameters that are subject to ML optimization.
We develop an RL controller that determines optimal control parameters based on the current environmental conditions, and find that the overall atom number achieved exceeds that of all other methods.  We specifically compare this RL controller to a  supervised regression model that maps the combined input space of control parameters and sensed environmental parameters to the number of atoms in the trap, and uses the model to find  optimal control parameters that maximize atom number for a given environmental state. 
Our results show that RL offers unique advantages to experimental control in high-dimensional systems, especially for its ability to react to drifts in the environmental conditions that have influence over the outcomes.

\section{Methods}

Ultracold quantum gas experiments operate, generally, by varying a large number of parameters over several seconds to achieve laser cooling, atom trapping, and additional evaporative cooling, with each cycle culminating in a destructive measurement of the atom's number and/or temperature. Here, we give a ML agent control over thirty parameters throughout the sequence, and supply the agent with measurements of thirty environmental parameters sensed during the previous cycle. Fundamentally, the question posed to the two agents we design is the same: given the current state of the environment, what is the ideal agent action (i.e., the settings of some controllable parameters) that maximizes the number of atoms imaged at the end of the cooling procedure?

\subsection{Experimental Apparatus}
\begin{figure*}[tb!]
    \includegraphics{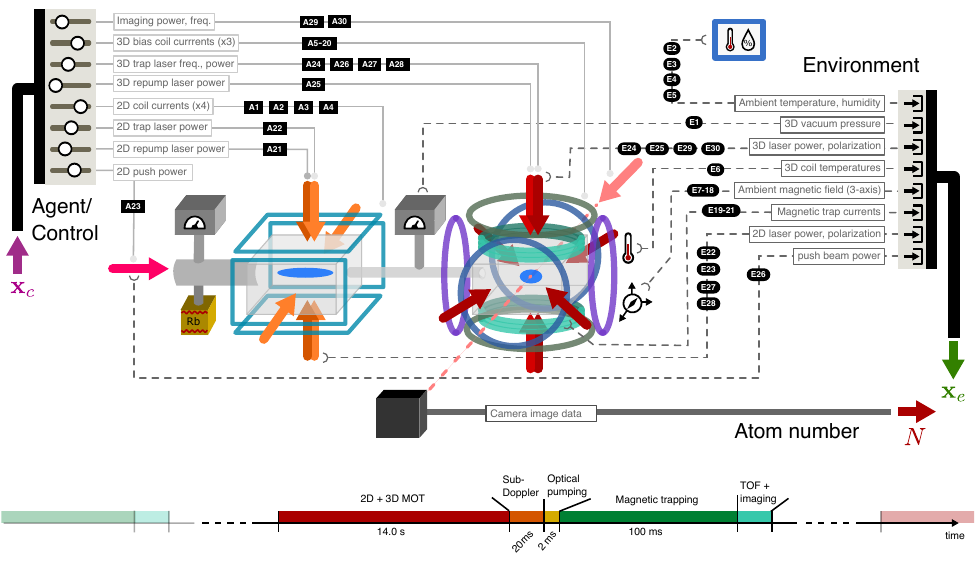}
    \caption{System schematic. {\bf Top:} A set $\mathbf{x}_c$ of agent-defined parameters (A$i$) control the experimental apparatus, while a set $\mathbf{x}_e$ of environmental parameters (E$i$) are measured during each experimental sequence.  The apparatus consists of an oven-fed 2D-MOT (with agent-controlled coil currents and laser powers), which supplies a 3D-MOT (with agent-controlled  laser powers and frequencies) via an agent-controlled push beam. At the 3D-MOT, several cooling steps precede the magnetic trapping of the atoms, agent-controlled bias magnetic field coils are used to optimize the environment throughout the sequence. The atom number $N$ is the reward and determined using data obtained by the camera, with agent-controlled imaging laser power and frequency.  {\bf Bottom:} The timing sequence for each experimental cycle, which is repeated after each destructive image of the atoms is recorded.} 
    \label{figa}
\end{figure*}

For this work, we use an experimental apparatus designed and built to create BECs of rubidium-87~\cite{Saglamyurek2021}. To achieve BECs with large, consistent numbers of atoms, we find that the large, consistent numbers of atoms at the magnetic-trapping (MT) stage of our sequence is key to success. Despite the best manual optimizations we apply, we observe long- and short-term drifts in the atom number, and suspect environmental changes are responsible for the instabilities.

Overall, our atom-cooling sequence begins with atoms under ultrahigh vacuum laser-cooled in a two-dimensional magneto-optical trap (2D-MOT), and then transferred to a separate chamber with a blue-detuned push beam to be laser-cooled in a three-dimensional (3D) MOT. Next, the atoms undergo motion-induced orientation cooling over a range of laser frequencies to cool them to sub-Doppler limit temperatures. Subsequently, the atoms are optically pumped into a magnetically trappable Zeeman state and transferred to a quadrupole magnetic field trap. Last, the MT is compressed by adiabatically ramping the  magnetic field gradient to significantly increase atomic density in preparation for forced evaporative cooling.  At this stage, we release the atoms in time-of-flight and resonantly absorption-image the cloud to determine the number of atoms. The temperature of the resulting ensemble is affected by the trap confinement. The ensemble trapped in the initial MT is $50$ $\mu$K, and increases to $450$ $\mu$K with compression of the MT.

Throughout each stage in the cooling process, we have several degrees of control. Some examples include: laser powers and frequencies controlled via acousto-optical modulators during the 2D- and 3D-MOT phases affect the cooling power; the position of the magnetic-trap centres may be moved via three sets of Helmholtz-configuration bias coils  to align with the intersections of cooling beams to reduce heating when transitioning from one preparation stage to the next; or the atomic density can be adjusted by changing the current of gradient coils during the magnetic trapping phase. In total, we assign  30 such parameters to be controllable by our ML agents (see \ref{sec:control-params} for full list).

To inform the ML agents' decision-making process, a monitoring system comprising a variety of sensors is implemented to capture and track the relevant external parameters throughout the preparation stages. These sensors enable us to construct a representation of the environment state to which the agent responds by selecting appropriate control parameters. Environmental parameters are read at the completion of each full  cycle, with the control parameters held at fixed values. This approach decouples the subsequent environmental state readings from the previously chosen control parameters. By doing so, the control parameter choices no longer affect the environment, thereby alleviating some of the complexity of this control problem. The environmental parameters monitored encompass a wide range of factors, including room temperature, room humidity, magnetic trapping coil temperature, laser powers, magnetic field strengths, vacuum pressures, and coil currents. Multiple sensors are strategically placed and configured to track a total of 30 environmental parameters (see \ref{sec:env-params} for full list). 

\subsection{Reinforcement-Learning-Based Agent Design}

 RL tackles the optimization problem within a Markov decision process (MDP) framework by selecting actions to maximize long-term cumulative rewards. In general, actions influence the subsequent state of the environment, requiring agents to consider the impact of their choices on future rewards. In our case, we  designed our system such that the observed outcomes are decoupled from previous control parameter actions, making it akin to a contextual bandit problem \cite{wang2005bandit}. Contextual bandits can be viewed as an edge case of MDPs, where episodes consist of only a single step, or alternatively the discount factor is zero~\cite{sutton2018reinforcement}. Despite the reduced complexity resulting from the decoupling, we opt for a more powerful RL approach due to the continuously varying environment, the high dimensionality of the action space, our lack of knowledge about the objective function's true form, the need for relatively quick generation of control parameter actions, and the potential for a partially observable environment leading to heteroscedasticity~\cite{le2005heteroscedastic}. 

\begin{figure*}[tb!]
     \includegraphics{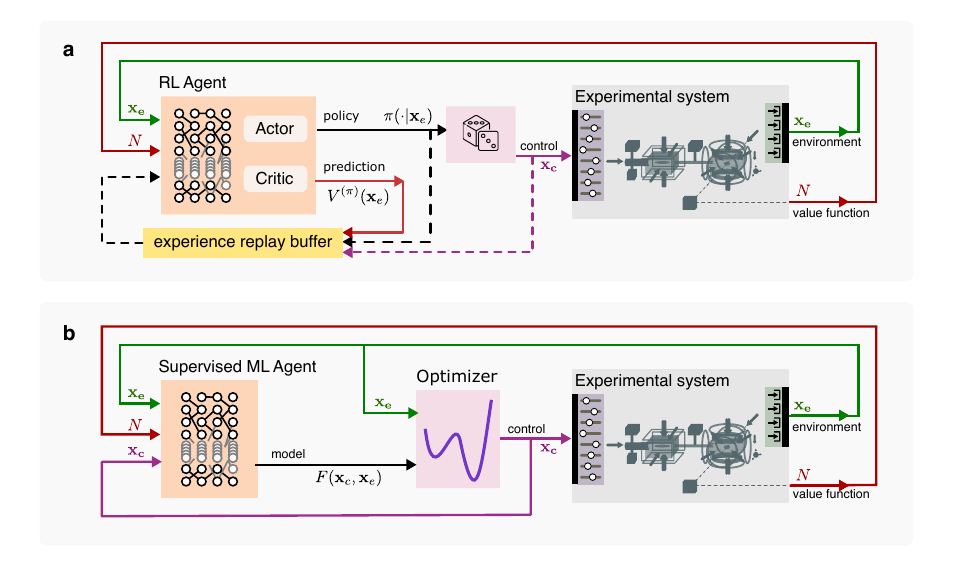}
    \caption{\textbf{a} Schematic diagram of the actor-critic agent’s live control loop. The agent leverages the sensed current environment as well as information about its own previous performance to output a continuous probabilistic policy, from which control parameter actions are drawn. The control parameters are implemented into our system, and the resulting cloud of cooled atoms is imaged. The weights and biases of the actor-critic network are then adjusted depending on the number of atoms in the cloud. This loop may then repeat. \textbf{b}  Schematic diagram of regression based agent's live control loop. The neural network regressor uses its bank of training data to makes a map between the concatenated control and environment parameter vectors, and the  numbers of atoms imaged. The network is then partially optimized to find the control parameters that maximize the outputted number of atoms, given the current environmental state. These maximizing control parameters are implemented into our system, and the resulting cloud of cooled atoms is imaged. With this additional experience tuple consisting of the new atom number and the control and environmental parameters that generated it, the network weights and biases are retrained and the loop may repeat. }
    \label{figb}
\end{figure*}

Unlike models trained to predict the desired output metric based on inputs, including the control parameters, the philosophy of RL is to explicitly train an agent to learn an optimal policy, from which control parameter actions may be drawn. Here, we build our agent as an actor-critic neural network. Actor-critic networks are well-suited for this problem, as they generalize naturally to continuous action spaces and learn stochastic policies which allow our optimizing agent some degree of exploration. We show a schematic of the control loop in figure \ref{figb}a.

The sole input to this network is the environmental state vector, $\textbf{x}_e$. The network output has two heads: the first head, known as the actor, seeks to output an optimal policy that chooses subsequent control parameter values, $\textbf{x}_c$. Considering the continuous 30-dimensional control parameter space in this problem, the actor is learning a 30-dimensional normal distribution. As such, the actor head consists of 60 output nodes, half of which are learning-control-parameter means, while the other half are the associated variances. The mean approximating nodes are bounded using hyperbolic-tangent activation functions, allowing the output values to be compatible with our system's hardware, while ensuring differentiability of the network through back-propagation. The variance-approximating nodes are followed by softplus activations, bounding the variances to positive numbers while remaining differentiable. The other head, known as the critic, consists of a single node. The critic learns a value function, which seeks to predict the resulting atom number while following the actor's policy, given the input environmental state. The internal architecture of our actor-critic agent consists of four hidden layers, each comprising 128 neurons. The network is initialized using the Glorot-Normal method~\cite{glorot2010understanding}, and network weight and bias updates are computed using the Adam algorithm~\cite{kingma2014adam}. To address the issue of exploding gradients during training, we employ Scaled Exponential Linear Unit (SELU) activation functions for each hidden layer~\cite{klambauer2017self}. Through experimentation with various activation functions, we found that SELU effectively mitigated the gradient instability. Furthermore, to prevent overfitting and promote sparsity in the network weights, we apply an $L_1$-norm penalty term as regularization. This regularization constraint not only helps prevent overfitting, but also has the potential to implicitly reduce the dimensionality of the problem~\cite{candes2008enhancing}.  

The network's critic node is optimized by iteratively adjusting the network's weights and biases by minimizing the squared error
\begin{align}
\label{critc}
\delta_{c} = (\log{(N)} - V_{}^{(\pi)}(\textbf{x}_e))^2,
\end{align}
where $N$ is the measured atom number and $V_{}^{(\pi)}(\textbf{x}_e)$ is the critic's value-function prediction, which  indicates the expected number of atoms (on a logarithmic scale) that the network predicts when control-parameter actions are chosen according to policy $\pi$, given the current environmental state $\textbf{x}_e$. 

The actor nodes are optimized according to the policy-gradient theorem~\cite{sutton1999policy}. For our case, in which subsequent environmental  parameters are decoupled from the previously chosen control parameters, the policy-gradient method results in practice as the iterative minimization of actor loss
\begin{align}
    \label{actor}
    \delta_{a} = - \log(\pi(\textbf{x}_c | \textbf{x}_e))  \cdot (\log{(N)} - V_{}^{(\pi)} (\textbf{x}_e)),
\end{align}
where $\pi(\textbf{x}_c | \textbf{x}_e)$ is the probability that control parameter action $\textbf{x}_c $ was chosen from the conditional probabilistic policy $\pi(\cdot| \textbf{x}_e)$.  The total non-$L_1$-penalized loss function is thus a weighted sum of the two constituent loss functions 
\begin{align}
\label{loss}
    L = \alpha_a \delta_a +  \alpha_c \delta_c,
\end{align}
where the weighting coefficients $\alpha_{a,c}$ are hyperparameters that can be thought of as different learning rates for the actor and critic. 

Despite the comprehensive sensor array we deploy, the complex nature of  atom cooling  leads us to hypothesize that the recorded environmental parameters may not fully capture the complete environmental state. This limitation, known as the issue of partial observability, means that the agent has access to only a subset of the pertinent information required for informed decision-making. Building upon prior research on autonomous curling robots~\cite{won2020adaptive}, we enhance the actor-critic network by incorporating temporal information from the agent's past performances. Specifically, we augment the environmental state with variables such as the previous atom number, the previous value function, and the probability of selecting the previous action based on the previous policy. Consequently, the agent can leverage not only the current environmental measurements but also its recent performance history when making decisions.

The vanilla policy gradient method used to train actor-critic models is generally an online algorithm, where the network weights and biases are updated after every new experience tuple collected. Here, an experience tuple consists of the environmental parameters, control parameters, and the atom number: $\left\{\mathbf{x}_e, \mathbf{x}_c, \log(N)\right\}$. To ensure gradient-calculation stability, we maintain a small experience-replay buffer~\cite{lin1992self} for performing batch stochastic gradient descent. Each new experience tuple is appended to the beginning of the buffer, while the oldest point is discarded. Additionally, to use the previously collected experiences from the supervised regression-based agent detailed below, we provide the actor-critic agent with a head start. We train the model on this bank of previously collected points as if they were being collected live, allowing the agent to benefit from this initial training data.

\subsection{Supervised-Regression-Based Agent Design}
As an alternative to RL, we implement a supervised ML technique, where a regression model maps a set of parameters to desired quantity, such as atom number; this class of ML is the one  used in several previous studies investigating ultracold atom preparation~\cite{wigley2016fast,tranter2018multiparameter,nakamura2019non,davletov2020machine,barker2020applying,xu2021maximizing,vendeiro2022machine,ma2023bayesian}. The universal function approximating models commonly used for cold atom optimization are Gaussian processes~\cite{seeger2004gaussian} and artificial neural networks~\cite{mehta2019high}. Here, we use a deep, feed-forward, densely connected artificial neural network to control cycle-to-cycle control-parameter variations, which vary in reaction to environmental factors. 

The input layer of the regression-based agent's neural network accepts a vector $\mathbf{x} = (\mathbf{x}_c, \mathbf{x}_e)$ that concatenates the control parameters and the environmental parameters, while the output layer predicts $\log{(N)}$. The internal architecture of the network consists of four hidden layers, each with 128 neurons. Gaussian-Error Linear-Unit activation functions follow each hidden layer~\cite{hendrycks2016gaussian}. The network weights and biases are initialized using the Glorot-normal method, and dynamically adjusted using the Adam algorithm. Performance evaluation is carried out using the Huber loss function~\cite{huber1992robust}. To prevent overfitting, training is terminated when the model's loss, validated on previously unseen data, no longer improves with subsequent epochs. It is important to note that, unlike the RL agent, we did not include additional elements in the environmental parameter vector to handle partial observability, such as information about the model's previous performance. Including these elements resulted in inferior control-parameter actions within our  architecture.   

The selection of control-parameter values in this scheme, as is common in similar experiments, relies on finding the set that maximizes the network's output. While neural networks may not in general possess explicit analytical invertibility, approximate optimization methods can be used, such as Limited-memory BFGS~\cite{byrd1995limited}, the gradient-free Nelder-Mead~\cite{nelder1965simplex} method, or probabilistic Bayesian optimization. In this experiment, we employ the Nelder-Mead algorithm, as it yielded the best results  given our time constraints. Although we explored Bayesian optimization with Gaussian processes as priors, the computational feasibility was limited due to the significant training time required for the Gaussian-process regressor: the computational complexity of the Gaussian-process regressor scales cubically with the number of points, making it impractical for optimizing the neural network within a reasonable timeframe, i.e., before significant changes in the environmental state occur.

We select the control-parameter set that maximizes the atom number, given a specific environmental state, by partially optimizing the model while holding the environmental parameters constant in the current configuration (Figure~\ref{figb}b). The control parameter set that maximizes the atom number, given an environmental parameter set, is 
\begin{align}
\label{regress update}
\mathbf{x}_{c}^{*} =  \underset{\mathbf{x}_c}{\mathrm{argmax}}\{F(\mathbf{x}_c, \mathbf{x}_e)\},
\end{align}
where $\mathbf{x}_e$ denotes the current environmental state, and $F(\textbf{x}_c, \textbf{x}_e)$ represents the network model approximating the atom number as a function of control and environmental parameters.

To train the network, data is initially collected  by sampling the control parameter space randomly. Once 1000 such points have been collected, we proceed to construct the training set using an iterative feedback loop. In this loop, the agent reads the current environmental state, adjusts the control parameters accordingly, creates an atomic ensemble, and measures the number of atoms. This experience tuple is added to the training data bank. By doing so, the network has an expanding training set to improve its model fitting. The feedback loop continues until a training set of $7000$ points is established. The size of this training data bank is  large compared to similar experiments~\cite{tranter2018multiparameter, barker2020applying}, which is necessary because our model incorporates environmental parameters that our agent cannot directly control, meaning that many iterations are required to obtain a representative sampling of the combined control and environment space.

\section{Results}

\subsection{Reinforcement-Learning-Based Agent}

The RL-based agent is initially trained on an offline-collected dataset and subsequently refined through live training using the learned probabilistic policy. The training set ultimately comprises $10~005$ experiences collected over a period of approximately three months. 

\begin{figure*}[tb!]
    \includegraphics{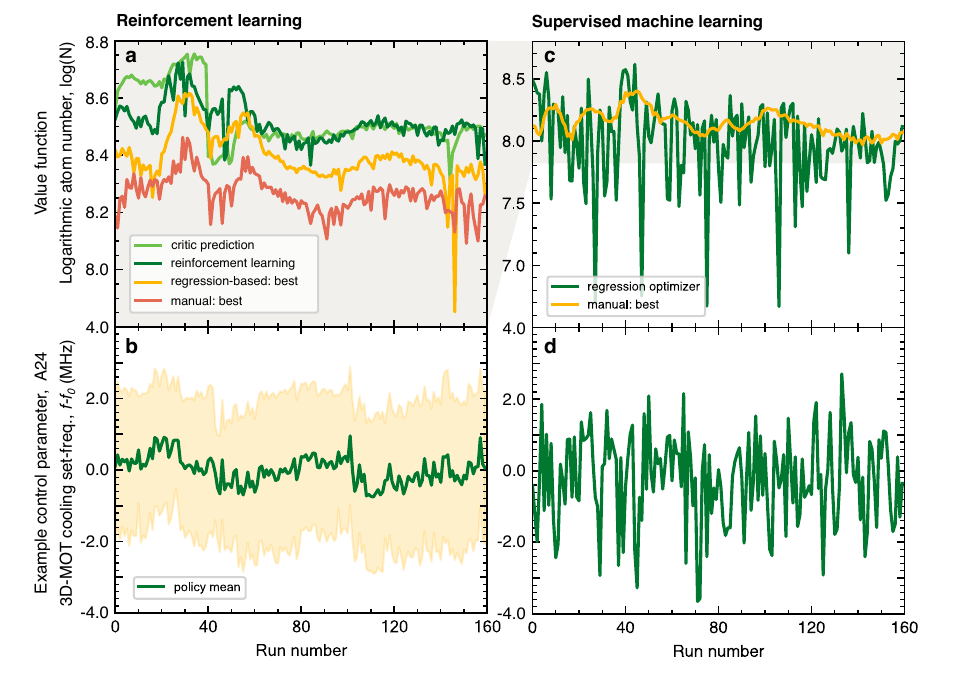}
    
    \caption{Machine learning performance. \textbf{a} In terms of the target atom-number value function, we compare  the reinforcement-learning actor-critic's live control of the control parameters (top, dark green) vs.\ fixing the control parameters at the best human-optimized set (red) and the best ever set from our regression-based agent (gold). Each run of these three approaches are performed one after another so that the system is subjected to approximately the same environment for all three approaches, i.e., the RL agent takes an action, followed by the fixed regression-agent derived action, followed by the human optimized action. Control parameters from the previous experimental run do not effect the current experiment, so the three approaches act independently. Note here that a run for a given approach constitutes applying the cooling sequence, followed by extracting the resulting number of atoms from the TOF image. This pattern then repeats 160 times.  The critic's prediction of the agent's performance is also shown (light green). \textbf{b} MOT laser cooling frequency (control parameter A24) policy, showcasing the mean value (green), and one standard deviation (gold). \textbf{c} Live performance of the supervised ML agent (green), noting that scale is expanded compared to the RL results, in light of the large fluctuations in the performance.  \textbf{d} Optimized MOT laser cooling frequency (control parameter A24) policy determined by the supervised ML agent.}
\label{fige}
\end{figure*}

To evaluate the performance of the agent, we assess its actions by implementing the control-parameter actions deterministically, i.e., solely from the policy's mean. For a comprehensive comparison, Figure \ref{fige}a shows the agent's actions with two alternative baselines: control parameters fixed at a human-optimized set, and fixed at the best set obtained from our regression-based agent. We assess the agent's performance over a period of 160 experimental runs. This period is long enough to assess the capabilities of the agent, while being small relative to the size of the training set, so that any change to agent's network weights will have a negligible effect on performance.  We find that the RL-based agent  generates control parameter actions that outperform human optimization and the regression-based approach, and does so consistently and autonomously cycle-to-cycle.  As an example of the control parameters used to generate this comparison, figure \ref{fige}b displays the MOT laser cooling frequency dimension of the multivariate policy. 

To further assess the effectiveness of our actor-critic network, we examine the predictive capabilities of the critic. Figure \ref{fige}a additionally presents a comparison between the atom numbers obtained by following the actor's policy and the critic's estimations. The critic's ability to accurately predict the atom numbers generated by the actor's policy serves as extra validation of the RL scheme, demonstrating the suitability of our environmental-state representation and network architecture.

\subsection{Supervised-Regression-Based Agent}

\begin{figure*}[tb!]
    \includegraphics{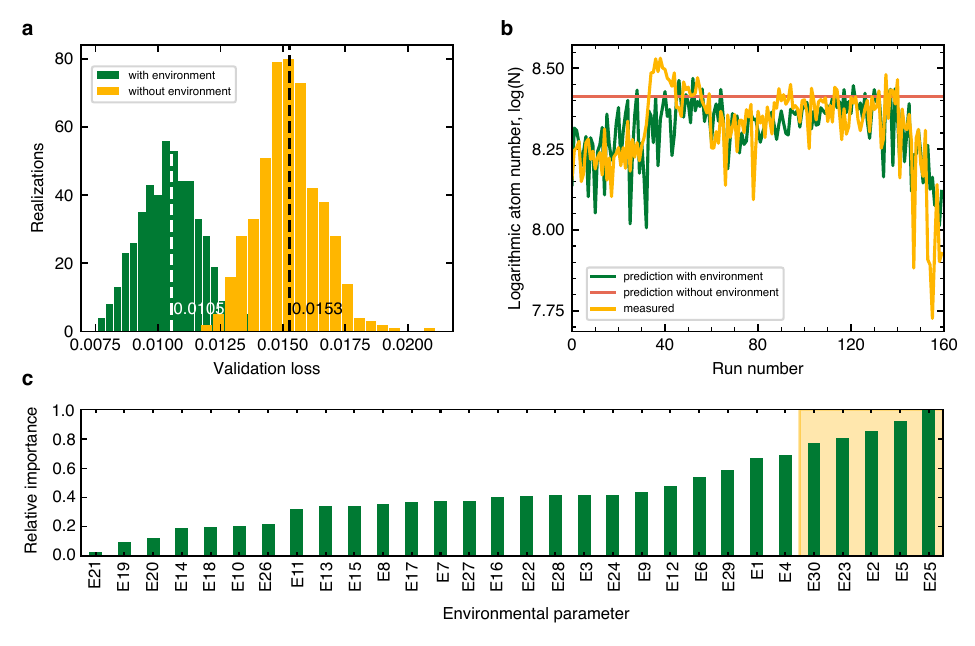}
    \caption{Supervised machine learning results.
    \textbf{a} Histograms of validation data loss, for models including (green) and excluding (gold) environmental parameters. The model is retrained 500 times with a different randomly selected validation set each time. \textbf{b} Model prediction (gold) of drifting atom numbers (green). Control parameters are fixed, such that the drift is driven solely by environmental changes. The model's prediction (red) is also shown when excluding the environment. \textbf{c} Importance of environmental parameters E$i$ (see Appendix for parameter names). The five most important parameters (highlighted gold) include the room temperature (E5) and humidity (E2), as well as the 3D-MOT cooling power (E25) and polarization (E30), and the 2D-MOT cooling power (E23).}
\label{figd}
\end{figure*}

We retrain this agent on the full bank of experiences, including those collected by the RL agent's exploration, thus resulting in a total data bank of $10~005$  experiences. This experience data bank was then randomly divided into training and validation sets using an 85:15 split. To ensure the robustness of this approach, we repeated the training process 500 times, each with different random partitions of the data. On average, we observed a significant reduction in our loss metric when incorporating environmental parameters into the model (Figure \ref{figd}a). 
We further evaluated the predictive power of the model on a separate set of testing points, where the control parameters were kept constant and atom-number fluctuations were driven by the  environment only. Our model successfully predicted (varying) atom numbers over periods as long as 2~hours, consisting of 160 iterations (Figure \ref{figd}b).

Our model successfully captures complex relationships between the control and environmental parameters, allowing it to accurately predict the corresponding number of atoms, more generally demonstrating that incorporating sensed environmental parameters into the model enhances its predictive performance.  

After training the regression-based agent and verifying its predictive capacity, we  evaluate its performance in generating control-parameter actions in response to different environmental conditions. The effectiveness of the agent is assessed on two primary metrics: achieving a high overall atom number and maintaining consistent and stable performance over time. To evaluate its live control capabilities, we grant the agent full control of the control parameters for a continuous period of two hours. For comparison, we also assess the performance by fixing the control parameters at the best configuration determined by human experts in our laboratory. 

Figure \ref{fige}c illustrates the results obtained from the agent's live control and the human expert's best set. While the agent occasionally achieves control parameter sets that surpass human performance, there remains a significant variance in the resulting atom numbers, which is consistent with observations from similar experiments. Over our 160 cycle showcase, the resulting $\log(N)$ generated by the agent has a variance of 0.14, compared to fixing the control parameters at the human expert's best set resulting in a $\log(N)$ variance of 0.01.   In figure \ref{fige}d, we show the agent's chosen MOT laser cooling frequency as an example of the large cycle-to-cycle variability of the control parameters. This variability limits the agent's suitability for live adaptive control and emphasizes the necessity of RL techniques to address this challenge.

Although  regression-based agents may be inadequate for live control of atom cooling procedures, their predictive capability makes them valuable for experimental design purposes. By leveraging this trained model, we can gain insights into the influence of different measured environmental parameters on the resulting number of atoms. This analysis helps experimental designers identify the critical components of the apparatus that require improvement and optimization. 
To estimate parameter importance, we use a feature permutation algorithm \cite{fisher2019all} (See \ref{sec:importance}), and  we present this metric for all parameters in figure \ref{figd}c.

\section{Discussion}
In a cold-atom experiment with a high-dimensional control-parameter space that is subject to a fluctuating environment, we find that optimizing control parameters with both a standard regression-based-ML protocol and a reinforcement-learning protocol result in larger numbers of atoms than achievable by our expert human optimizations.  However, only the RL algorithm behaved more consistently than the human optimization. Even in what is likely an only partially observable environment, where the RL agent is not given the capability to fully counteract all the environment driven fluctuations, it is still able to outperform all other baselines consistently.

The regression-based protocol exhibits strong predictive capability in estimating the atom number given the current control and environmental parameters. Using optimization, it can effectively suggest high-quality control parameter sets, although their variability limits their effectiveness for cycle-to-cycle optimization tasks. Nevertheless, regression-based methods remain valuable in scenarios where continuous control is required in the presence of an influential environment, as they enable the extraction of estimations regarding the importance of environmental parameters.

 With careful selection of the network's architecture and hyperparameters, the RL algorithms  consistently outperform human optimization and our regression-based agent in generating control parameter actions. It is worth noting that due to the large environmental- and control-parameter spaces, reliable simulators for generating large training datasets are not available. As a result, all iterations of the training loop were performed live on the physical apparatus. Despite this challenge, the RL agent consistently generates high-quality control parameter actions. This suggests that when control and optimization of an atom cooling apparatus is required, but large training datasets are not available, an RL policy-approximating agent could still allow for the consistent production of high quality ensembles.

Overall, while our findings suggest the applicability of RL actor-critic models in controlling the production of ultracold ensembles, we envision this as just the beginning of this burgeoning area of research. Additional computing power could allow more-expensive global optimization of neural network regression models, such as Bayesian optimization with Gaussian processes. This could enable regression-based approaches to achieve similar, or even superior, performance compared to RL. A high-performing supervised learning model, forming a trustworthy mapping between the combined space of control and environmental parameters to resulting atom number, would have the additional benefit of allowing experimentalists to see how control parameter sets predicted to be high-performing react to perturbations in the environmental parameters, thus giving an estimate of control parameter robustness.  Other forms of RL such as Q-learning \cite{watkins1992q, gu2016continuous} or Monte-Carlo policy gradients \cite{lazaric2007reinforcement} may, too, hold potential for producing interesting results when applied to control and optimization of ultracold atom experiments. Additionally, discretization of the environment space to enable the application of standard contextual bandit solving algorithms like $\epsilon$-greedy or gradient bandits is a potential avenue for investigation. Moreover, the incorporation of more complex reward schemes, including ensemble temperature, cooling duty-cycle duration, and cloud shape, among others, could further enhance the performance of such RL approaches. If stability is valued above all else, a reward scheme that is maximized when the resulting atom number is equal to a target atom number could be employed. With a reward scheme considering phase space density along with atom number, and giving the agent control over the details of the evaporation trajectory, our RL protocol would naturally be extended to control the production of BEC. By integrating these  ML approaches with advanced environmental-state monitoring, such as magnetometry via the Faraday effect \cite{budker2000sensitive} or even using a convolutional neural network to extract non-obvious environmental factors reflected in the TOF images \cite{kaming2021unsupervised,zhao2022observing}, and implementing supplementary hardware configurations like additional coils for magnetic-field compensation~\cite{li2022bi}, RL control could emerge as a crucial tool in the toolkit of atomic physics research, and in other similarly high-dimensional experimental systems.

\emph{Note added in preparation:} While preparing this manuscript, we were made aware of a preprint demonstrating reinforcement learning for a cold-atoms experiment~\cite{Reinschmidt2023-kh}.

\appendix

\section{Control Parameters}\label{sec:control-params}

\begin{table*}[b]
\centering
\caption{Agent controllable parameters\label{tab:AgentParams}parameters}
\begin{tabular}{l c l l}
\hline\hline
 Name    & Number   & Stage  & Implementation \\
\hline
2D coil 1                   & A1 &    MOT     & current  to 2D coil 1  \\
2D coil 2                   & A2 &    MOT     &   current  to 2D coil 2  \\
2D coil 3                   & A3 &    MOT     &   current  to 2D coil 3  \\
2D coil 4                   & A4 &    MOT     &   current  to 2D coil 4  \\
\hline
3D x-bias   & A5 &    MOT    &  current to x-axis bias coils \\
& A6 &  Sub-Doppler  \\
& A7 &  Optical pump \\
& A8 &  Magnetic trap 1   \\
& A9 &  Magnetic trap 2   \\
\hline
3D y-bias  & A10 &  MOT          & current to y-axis bias coils\\
& A11 &  Sub-Doppler\\  
& A12 &  Optical pump   \\
& A13 &  Magnetic trap 1    \\
& A14 &  Magnetic trap 2 \\
& A15 &  Imaging  \\
\hline
3D z-bias & A16 &  MOT & current to z-axis bias coils\\
& A17 &  Sub-Doppler  &  \\
& A18 &  Optical pump  & \\
& A19 &  Magnetic trap 1   &  \\
& A20 &  Magnetic trap 2  & \\
\hline
2D repump  & A21 &  MOT & laser power, via AOM \\
2D cooling                  & A22 &    MOT & laser power, via AOM  \\
2D push beam                & A23 &    MOT & laser power, via AOM  \\
3D frequency                & A24&    MOT   & laser frequency \\
3D repump  & A25&  MOT & laser power, via AOM \\
3D cooling                  & A26&    MOT  & laser power, via AOM \\
SD start freq.              & A27 & Sub-Doppler  & laser frequency \\
SD end freq.                & A28 & Sub-Doppler  & laser frequency  \\
OP power                    & A29 & Optical pump   & laser power, via AOM  \\
Imaging freq. &  A30   &  Imaging  & laser frequency \\
\hline
\hline
\end{tabular}
\label{control params}
\end{table*}

For this investigation, we allow both the ML agents to control 30 parameters in the cooling sequence, as listed in Table~\ref{tab:AgentParams}, and identified by the letter A and a number. These parameters all play an important role in the cooling process. 

In the 2D~MOT, the magnetic field configuration is set by 2D-MOT coil currents A1-A4, while the efficiency of the cooling and re-pumping is controlled by the powers of the 2D-MOT laser cooling beams, A21 and A22. Transfer from the 2D- to 3D-MOT can be optimized by changing the acceleration of the atoms via push-beam power A23. Once in the 3D-MOT, laser frequency and powers are set by A24-A26, and the zero-point of the quadropole magnetic field may be shifted with $x$-, $y$-, and $z$-bias coil currents, A5, A10, and A16. Optimal values of the magnetic fields, laser powers and frequencies allows efficient cooling of a large range of velocity classes via the optical molasses force, which depends on the intensities and detuning off resonance of the beams. 

During sub-Doppler cooling, a range of atomic velocity classes are addressed via the start and end points of our laser's frequency ramp, defined by A27 and A28, respectively. The bias magnetic fields  during this step, due to bias coil currents A6, A11, and A17, cancel out external fields. 

The initial magnetic trap is preempted by an optical pumping beam, whose power is set by A29. A quantization axis is defined by the magnetic field along $y$, set via coil current A12, while additional bias fields along $x$ set via coil currents A7 and A18, cancel background fields. With the atoms in the MT, bias magnetic fields controlled by coil currents A8, A13, and A19 shift the zero point of the trap. When compressing the trap, again the magnetic-gradient zero may be shifted this time with bias magnetic fields via coil currents A9, A14, and A20.

Finally, the falling atoms are illuminated with the imaging beam of frequency A30, while the imaging quantization axis is defined by a  magnetic field created by the $y$-bias coil current  A15.

\section{Environmental Parameters}\label{sec:env-params}

\begin{table*}[tb!]
\centering
\caption{Monitored environmental parameters\label{tab:enviro}}
\begin{tabular}{l c l}
\hline
\hline
Name & Number & Implementation \\
\hline
Vacuum pressure &E1& Ion pump pressure gauge\\
\hline

Room temperature  &E2& Thermistor \\
 &E3&  \\
\hline

Room humidity &E4& Capacitive humidity sensor,\\
&E5&  \\
\hline

Coil temperature &E6& Thermocouple \\ 
\hline

magnetic field x component & E7& Magnetometer\\
&E8& \\
&E9& \\
&E10&  \\

\hline
 magnetic field y component &E11& Magnetometer \\
&E12& \\
&E13&  \\
&E14&  \\

\hline
magnetic field z component&E15&Magnetometer \\
&E16& \\
&E17& \\
&E18& \\
\hline

Maximum MOT current & E19& current transducer\\
Maximum MT current &E20&  \\
MT current stability &E21& \\
\hline
2D repump beam power &E22& Photodiode\\
2D cooling beam power &E23&  \\
3D repump beam power &E24& \\
3D cooling beam power &E25&  \\
push beam power &E26&  \\
\hline

2D repump beam polarization &E27&Photodiode pair \&  \\
2D cooling beam polarization  &E28&  polarizing beamsplitter\\
3D repump beam polarization &E29& \\
3D cooling beam polarization &E30& \\
\hline
\hline
\end{tabular}
\label{ext params}
\end{table*}

In this work, we monitor 30 environmental parameters relevant to our cooling process, labelled with E and a number (Table~\ref{tab:enviro}). We monitor the majority of parameters at the end of each cooling cycle, which decouples the measured environmental parameters from the chosen control parameter values. Exceptions to this timing are E19-E21, which  are measured during different stages of the cooling cycle (MOT for E19, MT for E20, E21) via current transducer. To prevent interferences, we do not allow our agent to control the maximum current or the current stability of our MOT and MT, and in this way, these  parameters remain uncoupled from the control parameter actions of the agents.

Our environmental measurements include E1, which measures the vacuum pressure of the oven chamber, which houses  solid $\mathrm{^{87}Rb}$ used as a source, and is connected to the 2D-MOT chamber. The science chamber, where the 3D-MOT is located, is at too low a pressure to be measured by the ion pump. Higher vacuum pressure may result in decreased trap lifetimes and inefficient cooling procedures.

Two probes are placed in our laboratory, each measuring room temperature and humidity. One is placed next to our power supplies (E2 and E4), while the other is placed next to the 3D-MOT (E3 and E5). Temperature and humidity both may have effects on the cooling procedure by changing the air index of refraction, slightly misaligning optical elements, and performance of the diode lasers. Furthermore, a thermocouple measures the temperature of the main magnetic coil used both in the 3D-MOT and MT. Changes in the coil temperature would change the resistance of the circuit, and therefore could affect the performance of the magnetic traps. 

Using a vector magnetometer placed next to the 3D-MOT, we take multiple readings at the end of each cycle. Along the $x-$, $y-$, and $z-$axes respectively, we measure the magnetic field when the $x$-axis bias coils are active, alone, at a fixed current (E7, E11, E15), when  the $y$-axis bias coils are active alone at a fixed current (E8, E12, E16), when  the $z$-axis bias coils are active alone at a fixed current (E9, E12, E17), and when the 3D-MOT/MT coils are active at a fixed current (E10, E14, E18). We use such a configuration to detect unwanted external fields along all three Cartesian axes, as well as to detect potential issues with our power supplies.

We divert a small percentage of the 2D- and 3D-MOT's repump and cooling light for monitoring of the beam powers (E22-E25) and polarizations (E27-30). Drops in power and rotations of polarization would lead to inefficient cooling during both the MOT and sub-Doppler stages. The push-beam power is sensed after passing through the entire vacuum chamber and exiting a window near the 3D-MOT. An underpowered push beam fails to transfer the atoms optimally from the 2D-MOT to 3D-MOT, whereas if overpowered it can destroy the MOT's ability to effectively capture atoms.

\section{Parameter Importance Algorithm}\label{sec:importance}

The algorithm for estimating environmental parameter importance for our regression model is as follows:
\begin{enumerate}
    \item Train the model using an 85\% randomly selected subset of the data bank, incorporating all environmental parameters.
    \item Evaluate the Huber loss for the predictions on the remaining 15\% of the data, which serves as the validation set.
    \item Select an external parameter and randomly shuffle its values while keeping all other parameters unchanged. Re-evaluate the Huber loss on the same validation set to quantify the increase in the loss value, indicating the importance of the parameter.
    \item Repeat this process for each parameter column in the input array.
\end{enumerate}
We repeat the algorithm 50 times, selecting different validation sets at random in each iteration. 

\section{Estimating Atom Number From TOF Images}\label{sec:imaging}

After the compression of the MT, atoms are released from the trap and a destructive image is performed, allowing them a 10~ms TOF before they are illuminated with a $F = 2 \rightarrow F' = 3$ resonant beam. Three images are recorded by a CCD camera. The first image is timed to capture the atoms while they fall under gravity. The illuminating beam is partially absorbed by these atoms, casting a shadow on the camera. A second image is recorded after the atoms have fallen out of the field of view, with the absorbing beam  on: this measures the nominal intensity of the absorbing beam. A third image is recorded next, with neither beam nor atoms, and thus captures the background illumination. By subtracting the third image from the first two, and using Beer's extinction law along with the well-known absorption characteristics of the rubidium atoms~\cite{ketterle1999making}, we  measure the number of atoms at the end of each experimental sequence.
\begin{align}
    \label{number} 
    N = \sum_{num \ pixels}{\frac{A_{\rm pixel}}{\sigma_{\rm scs}}\ln{\frac{\textrm{CCD counts, no atoms}}{\textrm{CCD counts, with atoms}}}},
\end{align}
where $A_{\rm pixel}$ is the area of single pixel, $\sigma_{\rm scs}$ is the resonant scattering cross-section, and the sum is performed over every pixel in camera's  field-of-view.

\begin{figure*}[b!]
    \includegraphics[width=1\textwidth]{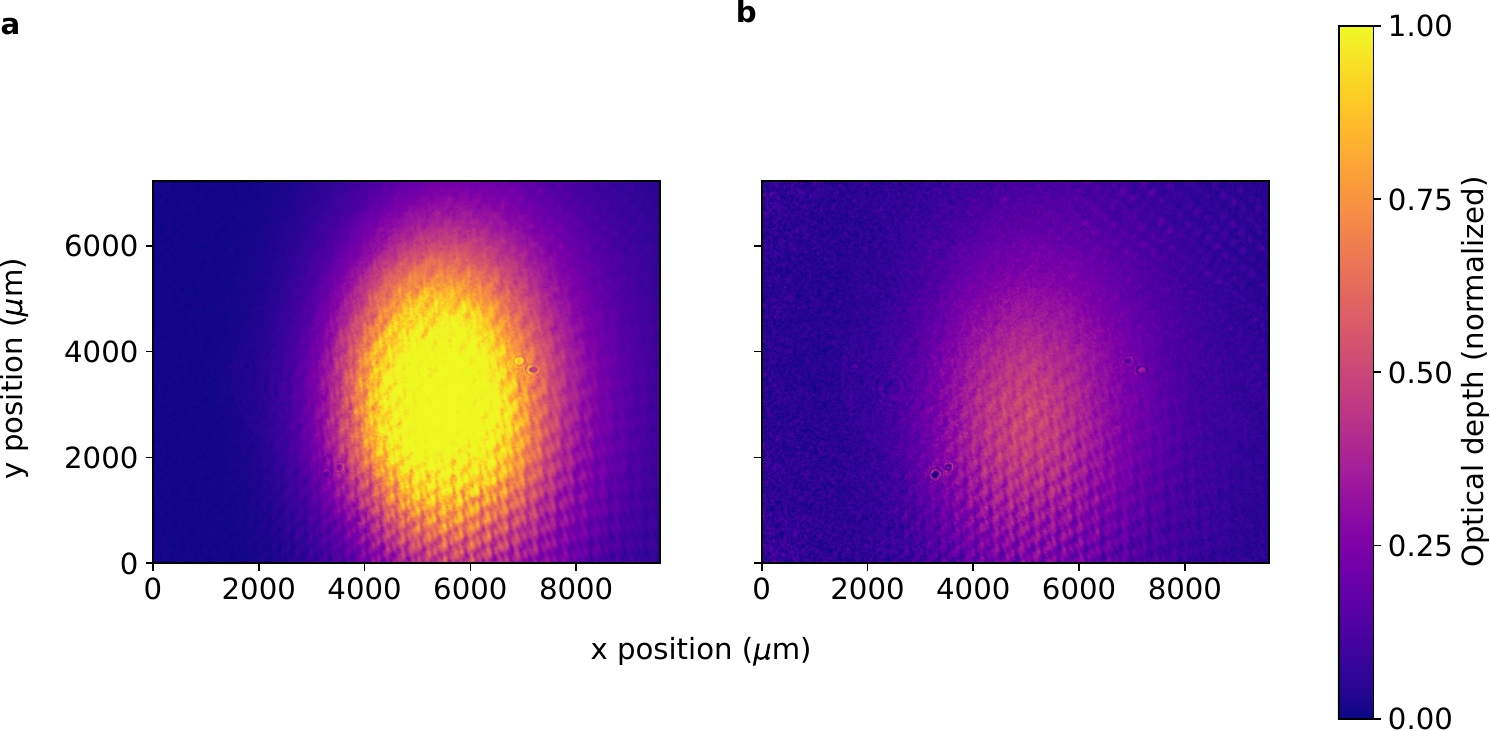}
    \caption{Processed TOF images. Each image is within a 648 pixel by 488 pixel field of view, where the pixel size is 14.8 $\mu$m. The colour scale shows the optical depth of the atoms in cloud, normalized to the maximum. \textbf{a} Sample ensemble generated by a well performing control parameter set. We estimate $2.3\times10^8$ atoms ($\log{N} = 8.37$) are present in this image. \textbf{b} Sample ensemble generated by a poor performing control parameter set. We estimate $3.37\times10^7$ ($\log{N} = 7.53$) atoms are present in this image.   }
\label{fig5}
\end{figure*}

We show samples of experimental TOF images in figure \ref{fig5}. These images are processed composites of the three raw images, allowing extraction of atom number estimates. 

\setcounter{section}{1}

\section*{Data availability}
The data supporting the findings of this study are available within the article and supplementary information. The main dataset collected for the RL agent is publicly available at GitHub \url{github.com/ultracoldYEG/Atom-Cooling-RL}, along with a sample script training the RL agent offline in preparation for live deployment. Additional data are available from the corresponding authors upon request.

\section*{Acknowledgements}
We would like to thank Abilmansur Zhumabekov for his expertise and fruitful discussions, and Logan W.\ Cooke, Benjamin D.\ Smith, and Taras Hrushevskyi for their work building and commissioning the apparatus.
This work was supported by the University of Alberta; the Natural Sciences and Engineering Research Council (NSERC), Canada (Grants No.\ RGPIN-2021-02884 and No.\ CREATE-495446-17); the Alberta Quantum Major Innovation Fund; Alberta Innovates; the Canada Foundation for Innovation; and the Canada Research Chairs (CRC) Program.
We gratefully acknowledge that this work was performed on Treaty 6 territory, and as researchers at the University of Alberta, we respect the histories, languages, and cultures of First Nations, Métis, Inuit, and all First Peoples of Canada, whose presence continues to enrich our vibrant community

\section*{Author information}
N.M., A.T., T.O., A.C., and L.J.L.\ built the physical experiment. N.M., A.T., T.O., and Z.F.A.\ designed the optimization schemes. N.M., A.T., T.O., and A.C.\ collected training data. N.M., A.T., and L.J.L.\ wrote the manuscript. All authors discussed the results and commented on the manuscript. The authors declare no competing interests. \\

\end{document}